# General Line Coordinates in 3D

Joshua Martinez
*Dept. of Computer Science*
*Central Washington University*
Ellensburg, WA 98926, USA
Joshua.Martinez@cwu.edu

Boris Kovalerchuk
*Dept. of Computer Science*
*Central Washington University*
Ellensburg, WA 98926, USA
BorisK@cwu.edu

*Abstract*—Interpretable interactive visual pattern discovery in lossless 3D visualization is a promising way to advance machine learning. It enables end users who are not data scientists to take control of the model development process as a self-service. It is conducted in 3D General Line Coordinates (GLC) visualization space, which preserves all n-D information in 3D. This paper presents a system which combines three types of GLC: Shifted Paired Coordinates (SPC), Shifted Tripled Coordinates (STC), and General Line Coordinates-Linear (GLC-L) for interactive visual pattern discovery. A transition from 2-D visualization to 3-D visualization allows for a more distinct visual pattern than in 2-D and it also allows for finding the best data viewing positions, which are not available in 2-D. It enables in-depth visual analysis of various class-specific data subsets comprehensible for end users in the original interpretable attributes. Controlling model overgeneralization by end users is an additional benefit of this approach.

## I. Introduction

3D graphics allows creation of detailed visuals on a variety of platforms, which can be a great asset to many applications as a well-functioning software for many industries, scientific computing and information visualization, video games, movies, television, advertising, and more.

The growing amount of data and computational power allows for processing of large amounts of data at high speed to solve multiple scientific, industrial, and healthcare problems. The problem is that the data now are multidimensional, and we cannot intuitively discover hidden patterns in these data. Therefore, to observe data and discover patterns in their full scope, lossless visualization methods that preserve all n-D data information are critical. Such capabilities will allow not only model developers, but also **end users/domain experts** who are, not data scientists, to develop, interpret and evaluate models, which is emphasized in the literature [7]. General Linear Coordinates (GLC) and Shifted Paired Coordinates (SPC) [1-3] are new areas in information visualization and machine learning that offer opportunities to achieve this goal.

General Line Coordinates in 3D (GLC-3D) for Visual Knowledge Discovery allows a user to see a separation of different classes of multidimensional data, and different unique rules which apply to the data without loss of multidimensional information in the 3-D visualization space. This is accomplished by displaying data in multiple shifted paired coordinates in combination with General Line Coordinates-Linear (GLC-L) both of which are special types of General Line Coordinates. This will allow a user to make informed visual decisions without explicitly using complex mathematical computations. Both SPC and GLC-L allow human-readable lossless representation of n-D points, which is not available in common dimension reduction methods like Principal Component Analysis. Prior systems with Shifted Paired Coordinates [3] and GLC-L [1] have so far been only in 2D.

The GLC-3D differs from them by adding the third dimension allowing a user to separate data, which before were hard to visually separate in 2-D visualization space. GLC-3D is created using the Unity engine [5] with C# code, a user interface, and DirectX's High Level Shading Language (HLSL) for shaders. GLC-3D is designed to feed the program with data from common machine learning repositories in csv format.

While an opportunity to decrease occlusion and overlap of the graphs by moving to 3-D is known, it is difficult to realize as it was demonstrated with parallel coordinates in 3D in the past. Selecting the right view projections is one challenge because the clarity of the data view heavily depends on both data and camera position. This paper demonstrates that decreasing occlusion and visual discrimination of classes is achievable in 3D GLC.

## II. Lossless SPC n-D data visualization in 2-D/3-D

### A. SPC in 2-D

The idea of SPC is illustrated in Fig. 1 from [3] for three classes of 4-D Iris data. Consider a 4-D point $\mathbf{p}=(p_1, p_2, p_3, p_4)$ and two 2-D points produced from it $(p_1, p_2)$ and $(p_3, p_4)$. We locate a pair $(p_1, p_2)$ in the Cartesian coordinates $(X_1, X_2)$ and a pair $(p_3, p_4)$ in the Cartesian coordinates $(X_3, X_4)$. These points are connected to form a polyline (directed graph). The example of these pairs is shown in Fig. 1 with the yellow line connecting them. Rectangle R`$_1$ indicates the rule of how the cases of Setosa class are separated from two other classes: If $(p_3,p_4)$ is in R`$_1$ then the case is in the Setosa class. Shifted Pair Coordinates have been implemented in the software [3, 6, 8] in 2D.

Next consider a more general situation of the rule discovery in SPC with a 6-D point $\mathbf{p}=(p_1, p_2, p_3, p_4, p_5, p_6)$ presented in Fig. 2 as three 2-D points $(p_1, p_2)$, $(p_3, p_4)$, $(p_5, p_6)$ in three shifted 2-D Cartesian coordinate systems. The same way as in Fig. 1 these points are connected to form a polyline (directed graph).

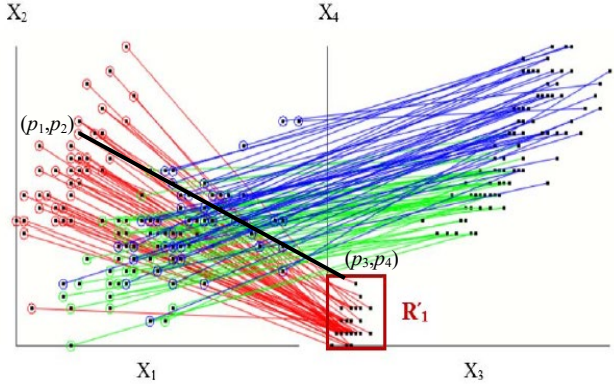

Fig. 1. Three classes of Iris data visualized in SPC in 2D implemented in SPCVIs [3] with rule R'$_1$ separating class Setosa (red) from other classes

Here we build rectangles R$_1$,R$_2$ and R$_3$ in the coordinate pairs (X$_1$,X$_2$), (X$_3$,X$_4$) and (X$_5$,X$_6$). In Fig. 2 in (X$_1$,X$_2$), the green rectangle R$_1$ is built around point ($p_1$, $p_2$) with using $\Delta_{1+}$ to get its right edge, $\Delta_{1-}$ to get its left edge, $\Delta_{2+}$ to get its top edge, and $\Delta_{2-}$ to get its bottom edge. Similarly, rectangles R$_2$ and R$_3$ are built for (X$_3$,X$_4$) and (X$_5$,X$_6$). These rectangles present a **logical interpretable rule** in SPC:

If ($p_1$, $p_2$) in R$_1$ & ($p_3$, $p_4$) in R$_2$ & ($p_5$, $p_6$) in R$_3$ then case **p** is in class 1.

This rule can be constructed directly in SPC visually as it was done in Fig. 1 or derived from some linear or non-linear discriminant function $f(\mathbf{p})$ [9].

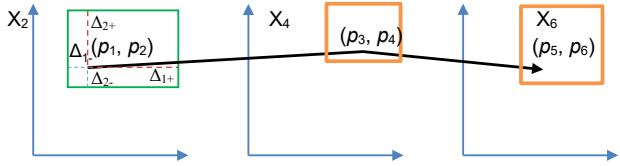

Fig. 2. Example of designing interpretable rule in Shifted Paired Coordinates.

B. *Shifted Tripled Coordinates (STP)*

There are two versions of SPC for 3-D space. The **first version** is when SPC are converted to **Shifted Tripled Coordinates (STP)**. It means that each n-D point $\mathbf{x}$=($x_1$, $x_2$, $x_3$, $x_4$, $x_5$, $x_6$,…,$x_{n-2}$,$x_{n-1}$,$x_n$) is split to triples ($x_1$,$x_2$,$x_3$), ($x_4$,$x_5$,$x_6$),…,($x_{n-2}$,$x_{n-1}$,$x_n$) and each triple is represented in its the 3-D cube as a 3-D point. For example, in the first cube the triple ($x_1$, $x_2$, $x_3$) is a 3-D point, where the values of $x_1$, $x_2$ and $x_3$ are mapped to X, Y and Z cube coordinates, respectively.

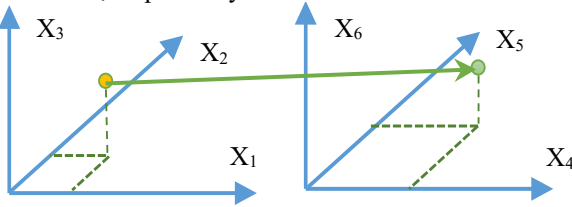

Fig. 3. Example of 6-D data in Shifted Tripled Coordinates (STP)

These cubes are shifted relative to each other. See Fig. 3. When dimension $n$ is not dividable by three then one or two attributes are copied to make $n$ dividable by three. To distinguish points of different classes the points and the connecting edges are colored according to the class color, e.g., Class 1 is red and Class 2 is blue.

C. *SPC in 3-D: Rectangular Rules and Discrimination Functions*

In the **second version** each cube encodes not three but **two attributes**. For instance, the first cube will encode a pair ($x_1$,$x_2$) which are mapped respectively to X and Y coordinates of the cube. The Z coordinate is used to represent the value of the **target attribute** $f$. It is especially useful for regression tasks and classification discrimination tasks, which use a threshold like, if $f(\mathbf{x})$>T then class 1 else class 2. Thus, it is still **Shifted Paired Coordinates but in 3-D space**.

In 3D it can be visualized in multiple different ways. One of the ways is using 2D visualization. It means that we still use **pairs** of coordinates to identify the first node of the SPC graph on the plane. We use the third dimension: (1) to connect nodes on the pairs of coordinates, (2) to distinguish nodes, which have the same pair of values, (3) to show classes. The advantage of using 3D visualization is the ability to spread the different graphs in 3D, which heavily overlap in 2D.

The use of the Z coordinate opens a new opportunity for SPC as we show below. Consider a discrimination rule R based on the linear function $f(\mathbf{x}) = a_1x_1 + a_2x_2 + … + a_nx_n$ as follows

   R: if $f(\mathbf{x}) \geq T$ then **x** is in class 1, else **x** is in class 2     (1)

The value of $f(\mathbf{p})$ is visualized as the **height** for the pair ($p_1$, $p_2$) in a 3-D SPC cube for this pair. Similarly, $f(\mathbf{p})$ value is visualized for ($p_3$, $p_4$) and ($p_5$, $p_6$) in their SPC cubes.

Then for a given point **p** from class 1 we compute the values of the linear function $f(\mathbf{p})$ and its parts $f_{12}(\mathbf{p}) = a_1p_1+a_2p_2$, $f_{34}(\mathbf{p}) = a_3p_3+a_4p_4$, and $f_{56}(\mathbf{p}) = a_5p_5 + +a_6p_6$, which includes only respective pairs of coordinates. We will call function $f_{12}(\mathbf{p})$ a **contribution** of ($p_1$,$p_2$) to $f(\mathbf{p})$. Respectively, we will call functions $f_{34}(\mathbf{p})$ and $f_{56}(\mathbf{p})$ **contributions** of ($p_3$,$p_4$) and ($p_5$,$p_6$) to $f(\mathbf{p})$.

Then $f_{12}(\mathbf{p})$ is visualized as a **white point** on the vertical line that shows $f(\mathbf{p})$ for the pair ($p_1$, $p_2$) in a 3-D visualization. The top of the vertical line for $f(\mathbf{p})$ is colored with the class color of **p**. See Figs. 4-5 with magenta for the class color. Similarly, $f_{34}(\mathbf{p})$ is visualized as the white point for ($p_3$, $p_4$). The greater height shows the greater **contribution** of the respective pairs. The same is done for ($p_5$, $p_6$). In general, the use of the function $f(\mathbf{x})$ allows analysis of the data of a **single class** without requiring a threshold of rule (1) to classify cases relative to alternative classes. In this way, each 2-D SPC plot is converted to a more informative 3-D plot.

Figs. 4-5 illustrate this idea for several iris 4-D cases in SPC-3D. Using the Z coordinate and connecting the lines at different heights allow these lines to be visibly more distinct than in the flat SPC like in Fig. 1. In general abilities to see data from multiple projections allows for a better differentiation of n-D points. The white rectangles in Figs. 4-5 are interactively adjustable to capture different sets of cases for a rule. The data

that are outside the rectangle are all grayed out and the connecting polylines are removed. The top points of vertical lines for $f(\mathbf{p})$ are colored with the class color (magenta) for each $\mathbf{p}$ and contributions of pairs $(p_1, p_2)$ and $(p_3, p_4)$ are shown as white dots. In this case, the contribution of $(p_1, p_2)$ is much greater than the contribution of $(p_3, p_4)$ for each $\mathbf{p}$.

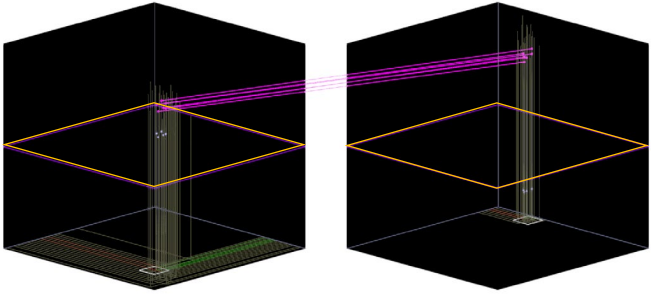

Fig. 4. Front view: several iris 4-D cases in SPC-3D and a rule shown with two white rectangles.

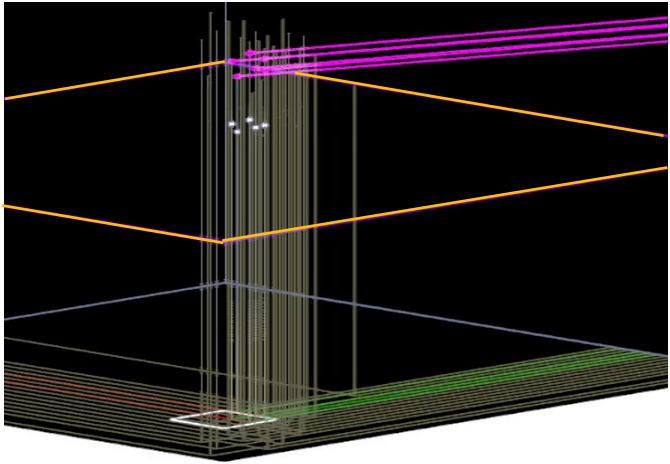

Fig. 5. Zoomed left cube with several iris 4-D cases in SPC-3D.

Figs. 4-5 show not only several Iris 4-D cases but also a horizontal **discrimination plane** for two classes. The height of this plane is the value of threshold $T$ from (1). All magenta lines that visualize 4-D points are above the plane, i.e., greater than $T$ and belong to class 1 in accordance with (1). This capability is absent in 2-D SPC.

Consider two different 4-D points: **a** and **b**. They can have the same values in the first pair and thus will be located at the same 2-D point on the horizontal $(X_1, X_2)$ plane in the left cube. In this situation **a** and **b** will differ in the right cube, e.g., **a**=(0.1,0.4,0.5,0.7) and **b**=(0.1,0.4,0.2,0.8). Their difference is captured in visualization in the right cube and in the points' height.

**Interpretable regression model**. Similarly, to building an interpretation model for a discrimination function we can build an interpretation model for the regression model. The accuracy of the interpretable rule for the regression model is controlled by the size of rectangles around point **x** in the respective SPC cubes. These rectangles need to be small enough as the antecedent of the logical rule R:

If $\mathbf{x} \in$ HB then $f_1 \leq f(\mathbf{x}) \leq f_2$ (2)

where

HB= $(a_{11} \leq x_1 \leq a_{l2})$ & $(a_{21} \leq x_2 \leq a_{22})$ & ... & $(a_{i1} \leq x_i \leq a_{i2})$
&...& $(a_{n1} \leq x_n \leq a_{n2})$ (3)

and $a_{ij}$ are limits of the sides of the respective rectangles and $f_1$ and $f_2$ are limits of the linear regression function $f$ in the HB. This is similar to a branch of the regression tree. From a mathematical viewpoint HB is a hyperblock, which generalizes an n-D hypercube allowing different lengths of its sides [4].

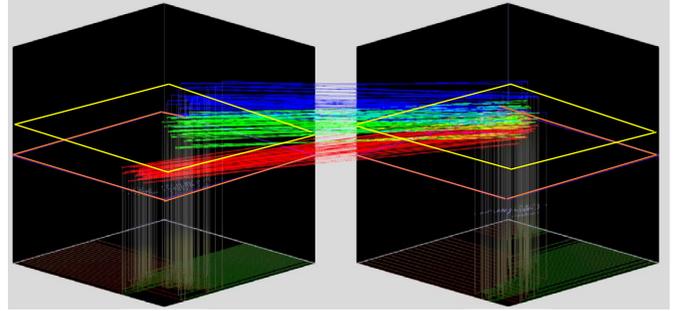

Fig. 6. Front view: Iris data with class-colored lines of 4-D points and white dots on vertical lines showing contribution of $(p_1,p_2)$ and $(p_1,p_2)$ to $f(\mathbf{p})$.

Figs. 6-7 illustrate this idea for both classification and regression tasks, where we built two horizontal planes at the levels $f_1$ and $f_2$. For classification these levels separate classes and for regression they estimate $f(\mathbf{x})$ by the interval $[f_1, f_2]$.

**Interactive visual approach.** The rules above are *not unique* because the intervals in (1)-(3) can be created differently and in multiple ways. To resolve this issue, we developed an *interactive visual approach* allowing a domain user to set up those intervals to make the rules practical for the domains. Figs. 6-18 show several views of all 150 Iris 4-D cases with contributions shown as white dots. These projections complement each other for a more complete data analysis.

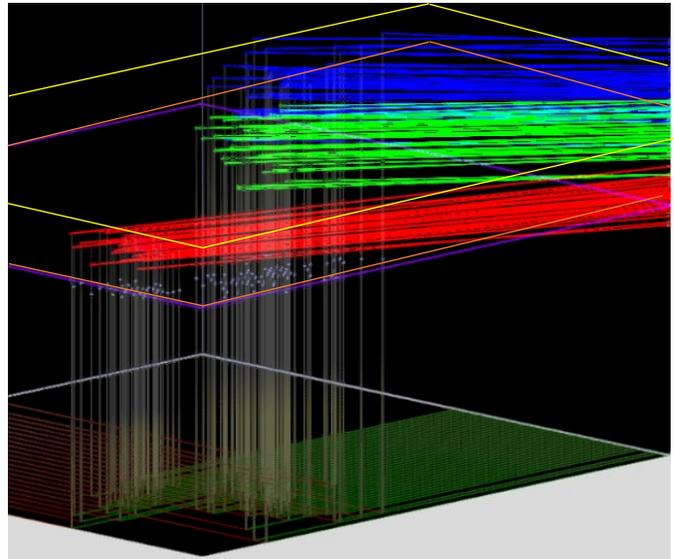

Fig. 7. Zoomed left cube: front of Iris data with vertical lines for $f(\mathbf{p})$ colored with the class color of each $\mathbf{p}$ and white contribution-level dots.

Fig. 7 is a zoomed left cube from Fig. 6. Fig. 8 is a top view, which is equivalent to 2-D SPC.

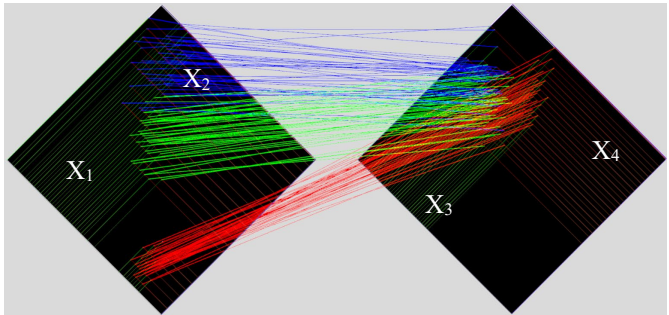

Fig. 8. Top view: Iris data in SPC-3D, which does not show this discrimination plane at all.

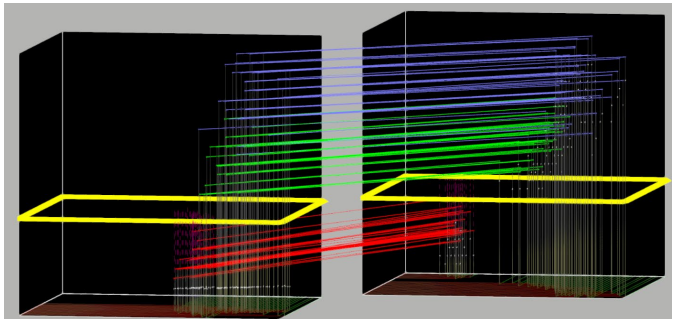

Fig. 9. Low front view: all Iris cases with a linear discrimination function/plane interatively found to separate blue and geent cases from red cases. Only red cases are below the yellow discrimination plane.

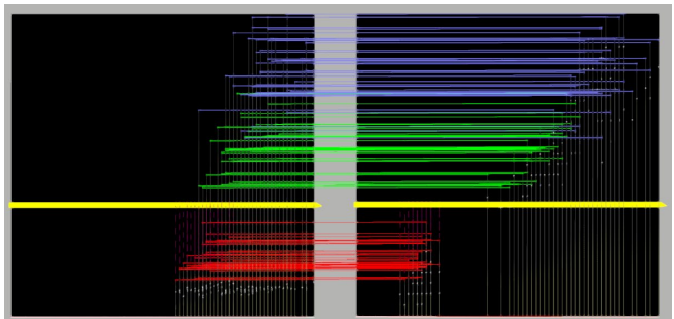

Fig. 10. Ortho left view of Fig. 9 with red cases below the yellow discrimination plane.

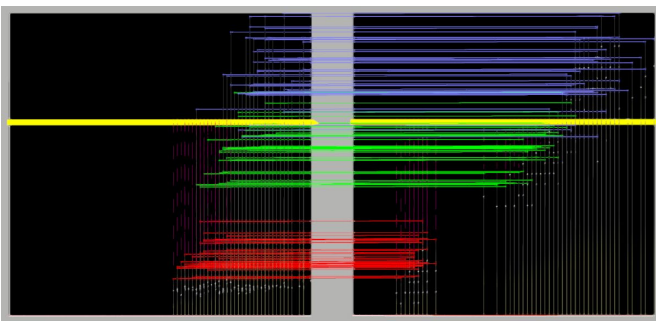

Fig. 11. Ortho left view: all Iris cases of three classes with a linear discrimination function interatively found to separate blue class from green and red classes.

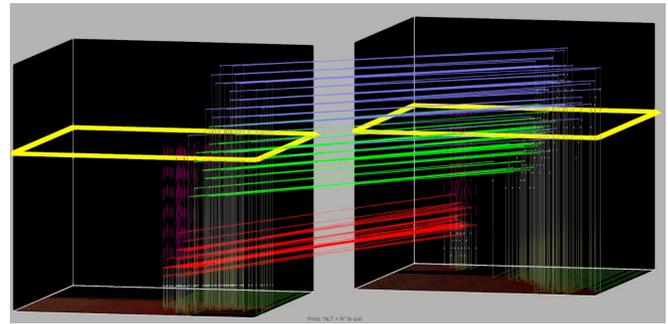

Fig. 12. Front view: all Iris cases of the three classes with a linear discrimination function interatively found to separate blue class from green and red classes.

Figs. 9-12 show different front projections of Iris data. Here Figs. 9-10 show separation of the red class from green and blue classes and figs. Figs. 11-12 show separation of red and green cases from the blue class.

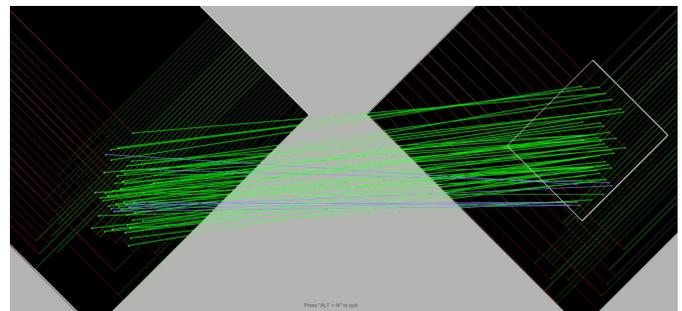

Fig. 13. Top view: all cases seelcted by a linear discrimination function interatively for the green class (including a few misclassfied blue cases). All these cases are also in the white rectangle on the right.

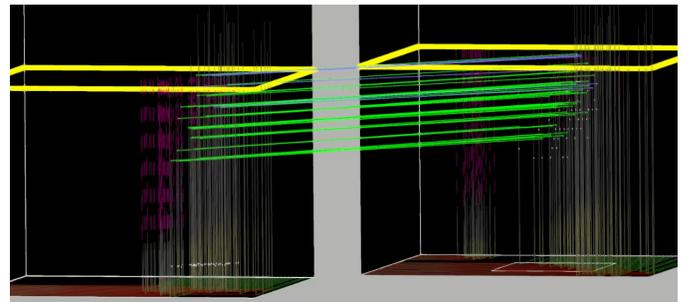

Fig. 14. Low front view: all cases seelcted by a linear discrimination function interatively for the green class (including a few misclassfied blue cases). All these cases are below the yellow discrimination plane.

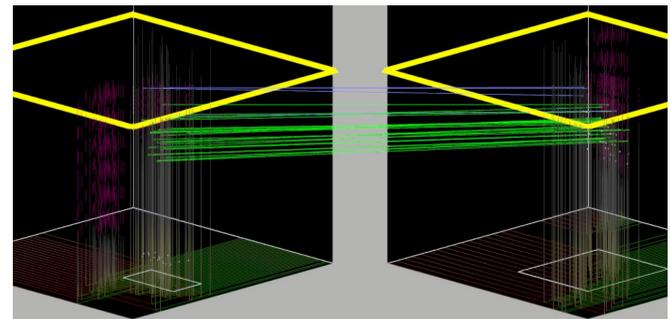

Fig. 15. Middle front view: all cases seelcted by a linear discrimination function interatively for the green class (including a few misclassfied blue cases). All these cases are below the yellow discrimination plane.

The separation of the blue class is not pure since a few blue cases are below the discrimination plane with figs.13-15 displaying this with different projections. Therefore, an interactive step was conducted to narrow the application of the separation plane to a subset of data by using a rectangle on the $(X_3,X_4)$ plane on the second cube. See Fig. 16. where a yellow rectangle excludes misclassified blue cases. The results are shown in Figs. 17-18, where blue cases are filtered out by decreasing the rectangle.

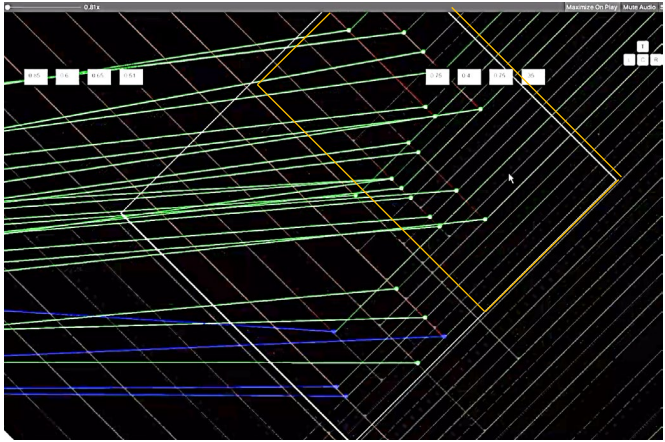

Fig. 16. Right cube top view: all cases seelcted by a linear discrimination function interatively for the green class. A yealow rectangle excludes misclassified blue cases.

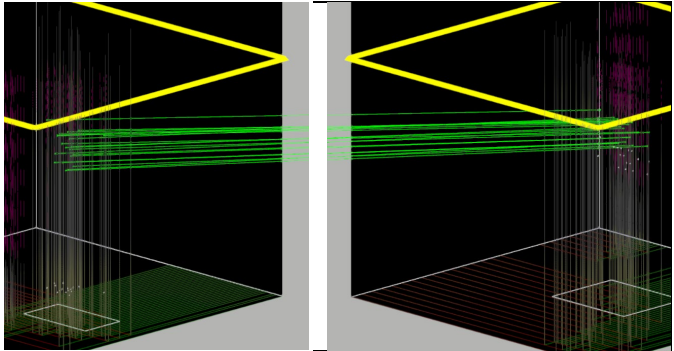

Fig. 17. Subset of Iris Virginica class in SPC-3D selected by two rules (the two white rectangles) and linear discrimination function (yellow plane).

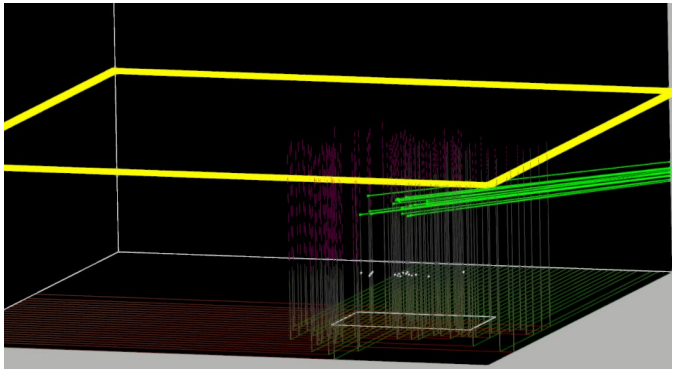

Fig. 18. Zoomed left cube from Fig. 19.

## III. GLC-L IN 3D

To explain the idea of the GLC-L method in 3D we start from the projection to the (Y,Z) plane shown in Fig. 19, which is equivalent to 2-D GLC-L presented in [1,13]. It shows two attributes $x_1$ and $x_2$ and the lengths of their projections on the vertical Z coordinate: 0.4. and 0.2. The total value, 0.6, is the value of a linear function $f(\mathbf{x})=a_1x_1+a_2x_2=0.4+02=0.6$. Here 0.4 is a result of projecting $x_1$ to Z using the angle $Q_1$, $0.4=\cos(Q_1)\cdot x_1$ and, similarly, $0.2=\cos(Q_2)\cdot x_2$. Thus, $f(\mathbf{x})= a_1x_1+a_2x_2= \cos(Q_1)\cdot x_1+ \cos(Q_2)\cdot x_2$ with respective coefficients $a_1= \cos(Q_1)$ and $a_2=\cos(Q_2)$.

Therefore, having any linear function $f(\mathbf{x})= a_1x_1+a_2x_2$ with coefficients $a_1$ and $a_2$ normalized to be in [-1,1] interval, we can find angles as follows: $Q_1$=arccos($a_1$) and $Q_2$=acrcos($a_2$). We can normalize coefficients using the *original function* $F(\mathbf{x})$ with coefficients $c_i$ as follows, $a_i = c_i/|c_{max}|$, where $|c_{max}|=\max_{i=1:n}|(c_i)|$. Here we rely on the following property

$F(\mathbf{x}) < T$ if and only if $f(\mathbf{x}) < T / |c_{max}|$.

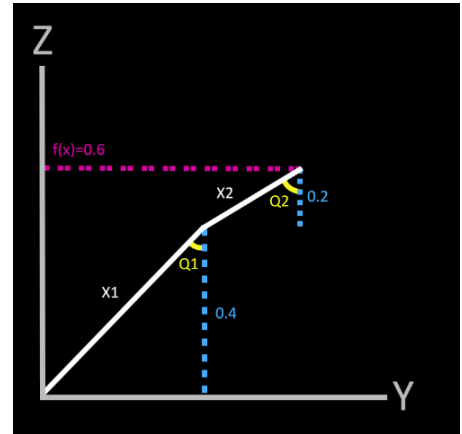

Fig. 19. GLC-L example left side view of (Y,Z).

While this example uses only $x_1$ and $x_2$ the formulas can be applied for any n-D point $\mathbf{x}=(x_1, x_2,…, x_n)$ and linear functions.

The next question is the location of a GLC-L polyline (graph) in 3D. So far we have shown how it is allocated on a single (X,Z) plane, but changing Y would allow multiple locations of these polylines in 3-D. Thus, in general, we can visualize a polyline in 3D using all three coordinates X,Y, and Z.

In a fully 3-D version, each $x_i$ value of an *n*-D point $\mathbf{x}$ can be located anywhere in 3-D, i.e., we can consider $x_i$ as a 3-D vector $(x_{i,X}, x_{i,Y}, x_{i,Z})$. The 3-D vector $(x_{i+1,X}, x_{i+1,Y}, x_{i+1,Z})$ for the next $x_{i+1}$ starts at the end of the vector for $x_i$. This version allows scaling the graphs using 3D space to decrease their overlap. It allows significant freedom in the location of the graphs. The only limitation is the angles to the Z coordinate, which represent the contributions of each $x_i$ to the target function *f*. Angles with X and Y coordinates can be assigned arbitrarily, e.g., randomly.

Another option is locating graphs on the vertical planes that go through the vertical line where the graph is originated on the (X,Y) plane. There are multiple such planes. The current implementation uses a plane that is parallel to the (Y,Z) plane. In contrast with SPC and STC all version of GLC-L in 3-D allow using the Z coordinate to represent the target attribute.

IV. JOINT SPC AND GLC-L IN 3-D

Above we described the implementation of GLC-L in 3D without link to the SPC. Fig. 20 shows how these two visualization methods are combined in 3D by embedding GLC-L in each SPC cube.

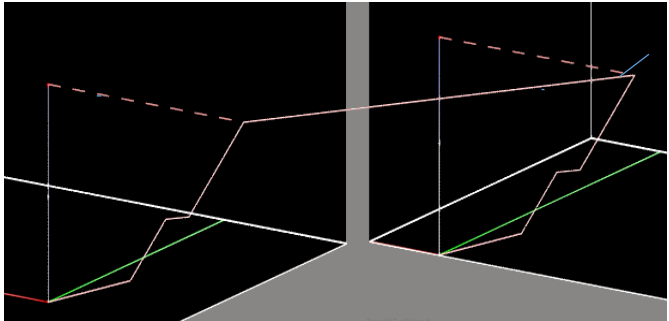
Fig. 20. Joint SPC and GLC-L in 3D

In the left cube GLC-L polyline starts at $(x_1, x_2)$ and in the right cube at $(x_3, x_4)$. These polylines are the same covering all 4 attributes $x_1$-$x_4$; thus, the lines have four segments. Here the GLC-L uses the third dimension Z. In contrast with traditional 2-D situation it spreads those GLC-L lines in 3D.

The program allows to control camera views to see the graph from different positions and to toggle several aspects of the visualization as shown in Fig. 21. Using the arrow keys, the user can manually rotate in 3D space. The ability to click the Left, Right, Center, and Top rotation buttons are available to reset the rotation or make larger rotations.

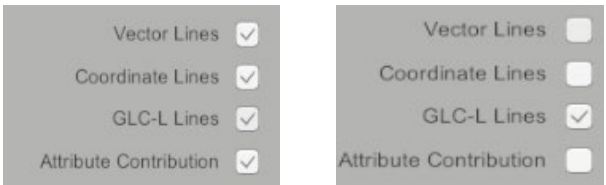
Fig. 21. Controls to toggle several aspects of the visualization.

Figs. 22-23 show a center camera view of Iris data in SPC combined with GLC-L in 3D, we will denote it as **GLC-3SL**. The $f(\mathbf{x})$ yellow plane is above the red cases in these figures.

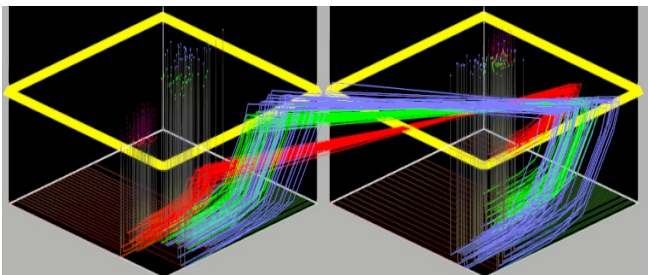
Fig. 22. Iris data in SPC combined with GLC-L in 3D

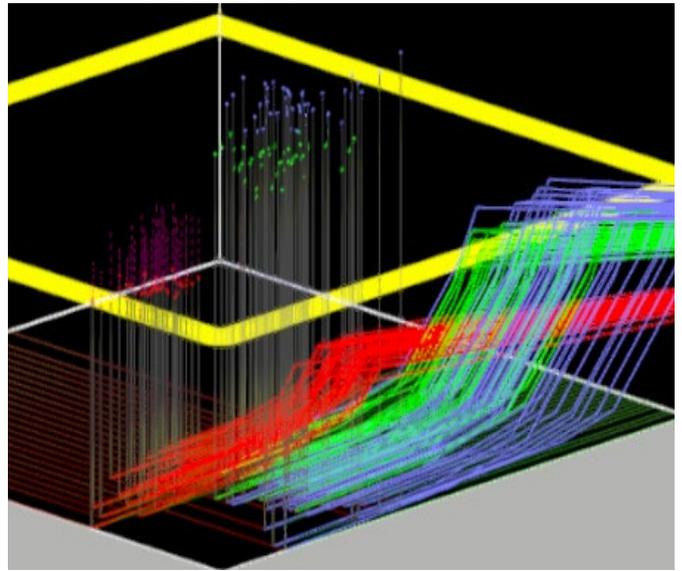

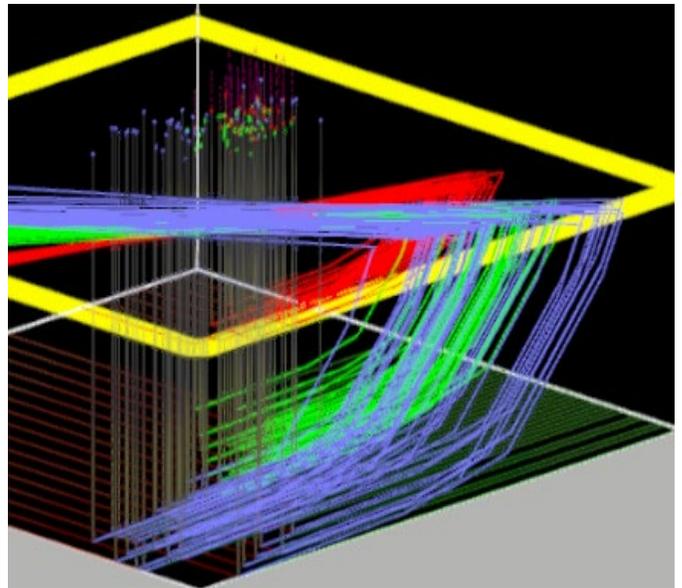
Fig. 23. Iris data zoomed from Fig. 21: top left cube, bottom: right cube.

V. REGRESSION IN SPC-3D

Above we described regression in SPC-3D using a pair of horizontal planes to approximate a given $f(\mathbf{x})$ value. It is applicable to all n-D points $\mathbf{x}$ with the same value $f(\mathbf{x})$. Below we present alternative ways expanding visualization options.
Fig. 24 shows alternative ways to represent a linear regression function using the geodesic levels and gradients of a linear function in 2-D in SPC for a 6-D point $\mathbf{x}=(x_1,x_2,x_3,x_4,x_5,x_6)$. While it simplifies visualization of the function it has difficulty to compare it with actual y values to observe errors due to lack of a 3rd dimension (the Z coordinate).

Another way to visualize the regression model in SPC for wider set of values of a linear function $f(\mathbf{x})$ is switching from 2D in Fig. 24 to **3D** in Fig. 25. This leads to a non-horizontal plane with different heights at different positions within the SPC cubes as shown in Fig. 25. It shows $f(\mathbf{x})$ for a given 6-D point $\mathbf{x}$ in SPC in the same ways as was used before in SPC-3D. In the left cube $f(\mathbf{x})$ is the height of the vertical line at point $(x_1,x_2)$. In the middle cube $f(\mathbf{x})$ is the height of the vertical line at point $(x_3,x_4)$ and in the right cube $f(\mathbf{x})$ is the height of the vertical line at point $(x_5,x_6)$.

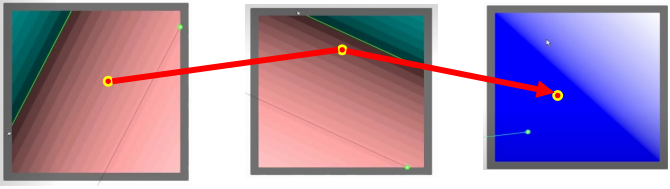

Fig. 24. 6-D point in SPC in 2-D with $f(\mathbf{x})$ captured by gradients.

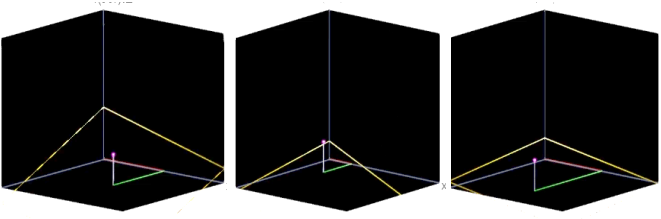

Fig. 25. Three planes with $f(\mathbf{x})$ value at positions $(x_1,x_2)$, $(x_3,x_4)$ and $(x_5,x_6)$ of the 6-D point $\mathbf{x}$. The height of the vertical line represents $f(\mathbf{x})$ value.

In the left cube we build a plane by computing $f(\mathbf{x})$ at the corners of left cube $(x_1,x_2)$: (0,0), (0,1), (1,0), (1,1) with fixed $(x_3, x_4, x_5, x_6)$ and similarly at the corner of the other cubes with fixed $(x_1, x_2)$. Fig. 25 shows these planes with value $f(\mathbf{x})$ is located on these planes. In the left cube it allows to vary values $(x_1, x_2)$ while keeping all $(x_3,x_4,x_5,x_6)$ constant in the visualization

$$f(x_1+\Delta_1, x_2+\Delta_2, x_3, x_4, x_5, x_6) \quad (1)$$

In this way a user can see the variability of $f$ relative to $x_1$, $x_2$ because the contribution of $x_3$-$x_6$ is fixed. Similarly, in other cubes, a user can see $f$ for different $(x_3,x_4)$, and $(x_5,x_6)$, respectively, with fixed contribution of remaining attributes,

$$f(x_1, x_2, x_3+\Delta_3, x_4+\Delta_4, x_5, x_6) \quad (2)$$
$$f(x_1, x_2, x_3, x_4, x_5+\Delta_5, x_6+\Delta_6) \quad (3)$$

If the actual **values** $y$ for a 6-D points $(x_1+\Delta_1, x_2+\Delta_2, x_3, x_4, x_5, x_6)$ are available then y values can be visualized in SPC at point $(x_1+\Delta_1, x_2+\Delta_2)$ in the left cube. It would allow the **differences** of $f(x_1+\Delta_1, x_2+\Delta_2, x_3, x_4, x_5, x_6)$ and y to be seen. The y values will be above or below this $f$ value, which is on the plane.

What is the benefit of such a visualization? We cannot visualize a function of four or more variables in 3D in the same way as we do with a function of 2-3 variables. Therefore, SPC-3D provides the ability to visualize $f$ in a new expanded but still limited way. In each cube it shows the behavior of $f(\mathbf{x})$ only of 2 attributes with all other attributes being fixed.

The views in cubes are **coordinated**, showing $f(\mathbf{x})$ for related 6-D points. It allows to a more **sensitive exploration** of $f(\mathbf{x})$ in the vicinity of $\mathbf{x}$. Changing $(x_1,x_2)$ in the first cube will change the planes in the other cubes too. The fact that the function is linear is not important; a non-linear function can be visualized in SPC-3D too.

## VI. RELATED WORK

In [10] the Immersive Analytics Toolkit, IATK, as a software package for Unity was introduced. It allows interactive authoring and exploration of data visualization in immersive environment. It includes assembling visualizations through a grammar of graphics that a user can configure in a GUI in addition to an API to create novel immersive visualization designs and interactions. IATK supports several million points at a usable frame rate.

In [11] 2D and 3D parallel coordinates are explored for temporal multivariate data. The results indicate that 3D parallel coordinates have higher usability, as measured with higher accuracy and faster response time as well as subjective ratings, compared to 2D.

In [12] Parallel Coordinate Plots (PCP) were explored in 3D using Virtual Reality (VR). The result displayed that it was easier to detect patterns in data using VR in comparison even to experts using the standard 2D PCP.

## VII. CONCLUSIONS

This paper contributes to machine learning via interpretable interactive visual pattern discovery in lossless General Line Coordinate space in 3D. This gives the end-user full control and exploration of the model development. We discussed challenges of algorithms in 3D and proposed the methods based on integration of Shifted Paired Coordinates and GLC-L methods for visualizing and discovering multidimensional patterns.

A major benefit of this 3-D technology is the ability to decrease the overlap of cases in visualizations and an expanding opportunity to find the best data viewing positions, which are not available in 2-D. It allows detailed visual analysis of individual data subsets of classes. This method is interpretable for the end-users since it is directly presented in the original attributes. Another advantage is the ability to control overgeneralization in models.

Future work will concentrate on enhancing SPC-3D and GLC-3SL for making: (1) classes visually well separated by finding projections where separation is visible, (2) non-linear scaling and non-orthogonal angles for better separation of classes, (3) visualization of higher dimensional data by adapting 2D serpent visualization [3] to 3D, (4) automated locations of areas where rectangles of cases of a class are concentrated (and counting the purity of cases within those rectangles), (5) genetic algorithms to find location of those rectangles, and (6) simplification of visualization.